\documentclass[a4paper]{article}
\usepackage{INTERSPEECH2022}
\usepackage{epsfig, fixmath, amsmath, amssymb, dsfont, xcolor,cite}
\usepackage{multirow}

\newcommand\sh[1]{\textcolor{black}{#1}} 
\newcommand\sr[1]{\textcolor{black}{#1}} 
\newcommand\ds[1]{\textcolor{black}{#1}} 

\title{Domain Agnostic Few-shot Learning for Speaker Verification}
\name{Seunghan Yang, Debasmit Das, Janghoon Cho, Hyoungwoo Park, Sungrack Yun}
\address{
 Qualcomm AI Research${}^{\dagger}$
  \thanks{ ${}^{\dagger}$ Qualcomm AI Research is an initiative of Qualcomm Technologies, Inc.}}
 \email{\{seunghan, debadas, janghoon, hwoopark, sungrack\}@qti.qualcomm.com}

\begin{document}

\maketitle
\begin{abstract}
Deep learning models for verification systems often fail to generalize to new users and new environments, even though they learn highly discriminative features.
To address this problem, we propose a few-shot domain generalization framework that learns to tackle distribution shift for new users and new domains.
Our framework consists of domain-specific and domain-aggregation networks, which are the experts on specific and combined domains, respectively.
By using these networks, we generate episodes that mimic the presence of both novel users and novel domains in the training phase to eventually produce better generalization.
To save memory, we reduce the number of domain-specific networks by clustering similar domains together.
Upon extensive evaluation on artificially generated noise domains, we can explicitly show generalization ability of our framework.
In addition, we apply our proposed methods to the existing competitive architecture on the standard benchmark, which shows further performance improvements.
\end{abstract}

\section{Introduction}
\label{sec:intro}

Deep learning models often fail to generalize to new domains and new classes, even though they produce highly discriminative representations.
Domain generalization \cite{li2018domain, li2018learning, li2019episodic, kang2020domain} and few-shot learning \cite{finn2017model, snell2017prototypical,das2022confess} have produced attractive solutions to address problems about unseen domains and classes, respectively.
However, only few studies \cite{guo2020broader, Tseng2020Cross-Domain} have focused on both problems, simultaneously.
In particular, identification systems always face both problems in the test phase while most approaches only consider the same set of categories and domains as they are trained upon.
Consequently, these closed-set models perform poorly on novel categories and novel domains.
\sh{This paper addresses this problem for speaker verification, which is the task of accepting or rejecting the claimed identity of a speaker test utterance based on few enrolled utterances.
}

Previous approaches \cite{heigold2016end, snyder2017deep, wan2018generalized, snyder2018xvec, li2018angular, heo2019end, liu2019large, ren2019triplet, yun2019end} on speaker verification train an embedding network using classification or metric learning loss on a dataset consisting of multiple speakers (users) in various domains. However, these methods do not generalize well to novel users and novel domains \sh{not involved during training}. There are several works~\cite{anand2019few, chung2020in, wang2019centroid} using few-shot learning approaches that generalize to new users. In training, they exploit prototypical networks~\cite{snell2017prototypical}, and apply an episodic learning scheme to mimic the test conditions of new users.
For classification, they use prototypes~\cite{anand2019few} or a class-centroid-based formulation~\cite{wang2019centroid} to compute the distance functions over all the classes.
These meta-learning based methods consider generalization on novel users but do not consider novel domains.
To achieve both objectives,
we propose a domain generalization component that mimics the presence of both novel users and novel domains in the episodic training phase.

{
In our framework, we consider two types of episode generations with domain-specific and domain-aggregation prototypical networks. Firstly, we generate the episodes consisting of the support and query set to mimic the cases of enrolling a novel user with few labeled samples (support set) and correctly classifying the test samples (query set). Here, the episodes are sampled from a specific domain and all domains for the domain-specific and domain-aggregation network, respectively. With the generated episodes and the prototypical loss function~\cite{snell2017prototypical, chung2020defence}, both networks are trained to adapt well to a new user. 
Secondly, we generate the episodes to mimic the cases of testing the network in a novel domain. 
Here, we consider a domain mismatched way: given a domain and its domain-specific network, the support and query sets are sampled from the other domain which is unseen to the specific network.
Then, the support and query sets are fed into the domain-specific and domain-aggregation networks, respectively.
With this episode generation and the prototypical loss function, we can train the domain-aggregation network with better domain generalization ability.
}

{
\sh{For the speaker verification task, the domains can be various environments such as background noises, recording devices, etc.}
However, modeling the domain-specific network for each domain may require much memory and computations. For the sake of efficiency, we consider a {\it pseudo} domain label approach where similar domains are clustered together using style features adopted in \cite{huang2017arbitrary, matsuura2020domain}.
}

\sr{We extensively evaluated our framework 
\ds{with two types of domain differences}: (a) artificially generated and (b) those intrinsically present in a dataset. Firstly, using the VoxCeleb1 dataset \cite{Nagrani17}, we artificially generated several domains with noise augmentations and split them into training and testing set.
In the experiments, we observed that the proposed framework shows better generalization ability for both new speakers and new domains in test dataset. Also, the pseudo domain labels are beneficial for efficient training and further performance boost. Secondly, our framework when applied on top of \cite{chung2020defence}, learns representations that can tackle
intrinsic domain differences in VoxCeleb2~\cite{chung2018voxceleb2}. The results also show performance improvements on the standard evaluation benchmark: VoxCeleb1, SITW~\cite{mclaren2016speakers} and CNCeleb~\cite{li2022cn}.}

\begin{figure*}[h]
	\epsfig{figure=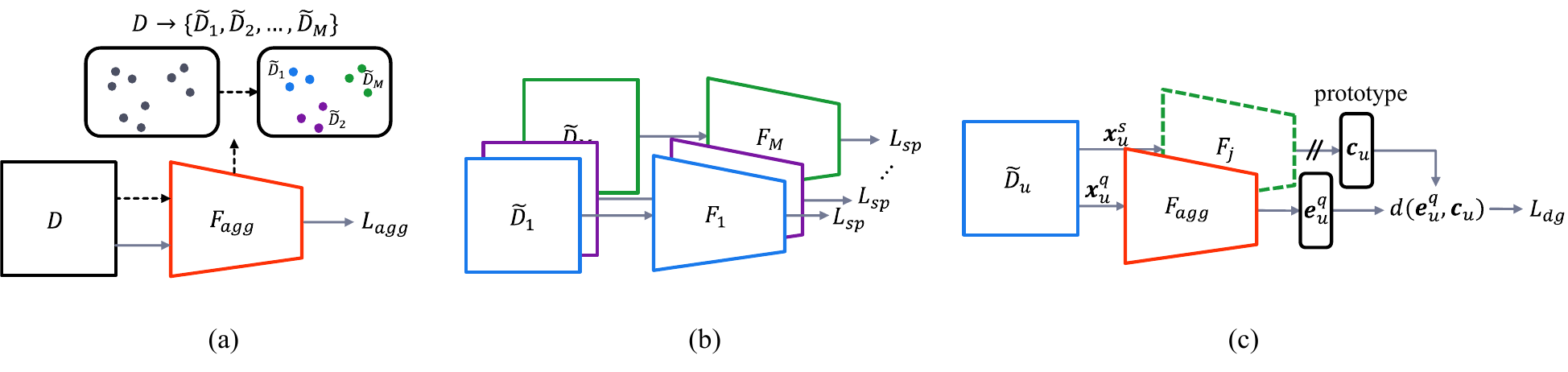,width=\linewidth}
	\vspace{-10pt}
	\caption{
	Our framework consists of: (a) Domain clustering: we assign pseudo domain labels to all samples using domain-discriminative features from the domain-aggregation network $F_{agg}$ pre-trained by $L_{agg}$; (b) Training domain-specific network: $F_{j}$ is trained on the corresponding source domain after clustering and re-assignment of domains; (c) Domain-agnostic few-shot learning: episodes are generated to mimic the few-shot test case for novel users and domain-mismatched case for novel domains.
	} 
\label{fig1}
\vspace{-10pt}
\end{figure*}

\section{Proposed Framework}
\subsection{Task Description}
In this paper, our task is speaker verification, which is a decision process that accepts or rejects an input utterance $\bf x$ by comparing the utterance with a trained reference speaker model ${\bf X}_{ref}$. Mathematically, the decision process is represented as:
\begin{eqnarray}
f({\bf X}_{ref}, {\bf x}) \underset{reject}{\overset{accept}{\gtrless}} \tau
\end{eqnarray}
where $f(\cdot, \cdot)$ is a similarity metric score between ${\bf X}_{ref}$ and $\bf x$. If the score is greater than a pre-defined threshold $\tau$, $\bf x$ is accepted as the utterance of the reference speaker; otherwise, $\bf x$ is presumed to be from an impostor. Our goal is to train a speaker verification model that generalizes to novel users and domains.


\subsection{Problem Setting and Notation}

We have a source dataset $D=\{D_{1}, D_{2}, … , D_{N}\}$ consisting of multiple speakers from $N$ domains, where $D_{j}$ indicates the $j^{th}$ domain. Each domain contains utterances and their corresponding speaker labels such that $D_{j}=\{({\bf x}_{j}^{i}, y_{j}^{i})\}_{i=1}^{n_j}$, where $n_j$ is the number of samples in the $j^{th}$ domain.
We use the source data to train an embedding network $F_{agg}$ such that it generalizes well to novel target users in novel testing domains $D_{*}$ with different characteristics from the source domains.
Note that it is different from data augmentation techniques that use noise and reverberation to enforce test domains to be included in the training domains.
We denote a domain-specific network and a domain-aggregation network by $F_{j}$ and $F_{agg}$, respectively. $F_{j}$ specialized in $j^{th}$ domain is learned so that $F_{agg}$ learned with all domains can extract domain agnostic features even for novel domain data. Fig.~\ref{fig1} shows our overall framework.

\subsection{Domain Clustering}
\label{sec:dc}
Our learning scheme requires $N$ domain-specific networks that consume lots of memory in training, which we can reduce by 
modifying the original source dataset $D$ with pseudo domain labels by clustering.
We extract domain-discriminative features and cluster them to assign the same pseudo domain labels to the samples of similar domains.
For domain-discriminative features, we exploit the features proposed in style transfer algorithms for different image domains~\cite{huang2017arbitrary, jung2020arbitrary, matsuura2020domain} that are recently shown applicable to audio domains~\cite{kim2021domain, kim2021towards, kim2021qti}.
In these methods, the mean and the standard deviation that are calculated across the spatial dimensions represent the input style.
We obtain the domain-discriminative feature of an input ${\bf x}$ by stacking these feature statistics as follows:
\begin{equation*}
\{ \mu (\phi _{agg, 1}({\bf x})), \sigma(\phi _{agg, 1}({\bf x})), ..., \mu (\phi _{agg, l}({\bf x})), \sigma(\phi _{agg, l}({\bf x})) \}
\end{equation*}
where $\phi _{agg, l}$ indicates the output of the $l^{th}$ layer in $F_{agg}$.
Before extracting features, we first train $F_{agg}$ with the source domain dataset $D$.
Then, with the extracted domain-discriminative features, we perform k-means clustering \cite{macqueen1967some} to obtain $M$ clusters and assign the pseudo domain label to each sample.
This is illustrated in Fig. \ref{fig1}(a).
Accordingly, we obtain a modified source dataset $D=\{{\widetilde D}_{1}, {\widetilde D}_{2}, ..., {\widetilde D}_{M}\}$ with pseudo domain labels, and $M \leq N$. 
In contrast to \cite{matsuura2020domain}, our objective of assigning pseudo domain labels is to increase memory efficiency by clustering ambiguous domains and reducing the number of domain-specific networks.

\subsection{Episodic Training}
\label{sec:epi}
In this step, we describe our two types of episode generations to mimic the test cases of the novel user and novel domain. First, we split the dataset $D$ into the support set $D^{s}=\{({\bf x}^{i}, y^{i})\}^{CK}_{i=1}$ and the query set $D^{q}$ to mimic the few-shot test case for novel users, which is a $C$-way $K$-shot problem. Here, way and shot stand for the number of speakers and samples per speaker in each episode, respectively. We use the prototypical network \cite{snell2017prototypical} to be trained in an episodic manner such that it computes speaker prototypes with the support set and forces the query set to minimize the distance from the corresponding prototypes.
\sh{Note that our algorithm can be adapted to any prototype-based loss functions, {\it e.g.}, angular prototypical loss~\cite{chung2020defence}.}

As illustrated in Fig.~\ref{fig1}(b), the domain-specific network $F_j$ is trained using episodes generated from ${\widetilde D}_j$, and prototype of speaker $k$, ${\bf c}_k$, is obtained by averaging over embedding vectors in $j^{th}$ domain support set:
\begin{equation}
\label{prototype}
{\bf c}_{k} = \frac{1}{K}\sum_{\{({\bf x}_{j}^{i}, y_{j}^{i})\}\in {\widetilde D}_{j}^{s}} F_{j}({\bf x}_{j}^{i})\mathds{1}(y_{j}^{i}=k)
\end{equation}
where $\mathds{1}(\cdot)$ is the indicator function returning $1$ for true statements and $0$ otherwise. The episodic training loss $L_{sp}$ for the domain-specific network is defined as follows:
\begin{flalign}
  &&~ &L_{sp} = \sum_{\{({\bf x}_{j}^{i}, y_{j}^{i})\}\in {\widetilde D}_{j}^{q}} -\log(p(y=y_{j}^{i}\mid {\bf x}_{j}^{i}))& \label{loss_sp}\\
  \text{where} 
  &&~ &p(y=y_{j}^{i}\mid {\bf x}_{j}^{i}) = \frac{\exp(-d(F_{j}({\bf x}_{j}^{i}), {\bf c}_{y_{j}^{i}}))}{\sum_{k'}\exp(-d(F_{j}({\bf x}_{j}^{i}), {\bf c}_{k'}))},& \nonumber
\end{flalign}
and $d(\cdot)$ is the Euclidean distance. 
Similarly, the domain-aggregation network $F_{agg}$ is trained by the loss function $L_{agg}$, which is the same with $L_{sp}$ except that the support sets and the query sets are sampled from all the source domains $D$.

The second type of episode generation and training is illustrated in Fig. \ref{fig1}(c). 
Our objective is to train $F_{agg}$ to extract discriminative features even for the novel domain data.
Hence, we construct the support set ${\widetilde D}_u^s$ and query set ${\widetilde D}_u^q$ from $u^{th}$ domain and mimic the test case of domain generalization exploiting $F_j$ (domain-mismatch).
With the domain-mismatch condition, {\it i.e.}, $j \neq u$,
\sh{unseen domain} prototypes ${\bf c}_{k}$ are calculated by Eq. \ref{prototype} using $F_{j}$ on the novel domain set ${\widetilde D}_{u}^{s}$, and embedding vectors of ${\widetilde D}_{u}^{q}$ are extracted from $F_{agg}$. The loss $L_{dg}$ for few-shot domain generalization episodes is as follows:
\begin{flalign}
  &&~ &L_{dg} = \sum_{\{({\bf x}_{u}^{i}, y_{u}^{i})\}\in {\widetilde D}_{u}^{q}} -\log(p(y=y_{u}^{i}\mid {\bf x}_{u}^{i}))&\\
  \text{where} 
  &&~ &p(y=y_{u}^{i}\mid {\bf x}_{u}^{i}) = \frac{\exp(-d(F_{agg}({\bf x}_{u}^{i}), {\bf c}_{y_{u}^{i}}))}{\sum_{k'}\exp(-d(F_{agg}({\bf x}_{u}^{i}), {\bf c}_{k'}))}.& \nonumber
\end{flalign}
Since the $u^{th}$ domain data is novel to $F_{j}$, $F_{agg}$ learns to map embedding vectors in a domain-agnostic way. Here, only $F_{agg}$ is updated by $L_{dg}$ since we need to keep ${\widetilde D}_u$ novel to $F_j$ during training. Our training procedure is summarized as follows: 1) Train $F_{agg}$ using $L_{agg}$ and extract domain-discriminative features, 2) Perform clustering on extracted features and assign pseudo domain labels: this leads to $M$ domains from $N$ input domains, 3) Generate $M$ domain-specific networks and train them using Eq. \ref{loss_sp}, 4) Simultaneously, train $F_{agg}$ using the combined loss $L_{agg} + \lambda_{dg}L_{dg}$ where $\lambda_{dg}$ is a balancing parameter.
\vspace{-10pt}

\section{Experimental Results}

\subsection{Experiment Setup}
{\bf Dataset.} First, to analyze generalization ability of our framework, \sh{we designed the evaluation experiment of simulating domain differences by artificially producing data augmentations.}
We selected 1,088 speakers in VoxCeleb1~\cite{Nagrani17}, consisting of 1,000 speakers for training and 88 for testing.
Each speaker has 45 utterances, and each utterance was randomly cropped into 2-sec-long for the use-cases of short utterance enrollment and testing. 
\sh{To explicitly generate several domains, we augmented domains by mixing noises.
Then, we split them into train and test sets to simulate the test condition where data are recorded in novel domains.}
The source dataset consists of 4 domains: original set (clean) and three noisy domains which were augmented by synthesizing the clean with babble, car, and music noises (0dB SNR).
Each of the novel speaker's data in the target domain consists of 5 utterances for enrollment and 35 utterances for testing. The target domain consists of 7 domains: original set (clean) and 6 noisy domains which were augmented by synthesizing the clean with babble, car, music, office, ambient room, and typing noises (0dB SNR). Thus, our model is evaluated on 4 in-domain sets ({\it i.e.} clean, babble, car, music) and 3 out-domain sets ({\it i.e.} office, ambient room, and typing).
Additionally, to analyze models in diverse environments, we augmented VoxCeleb1 with various noise types and SNR values for source and target domains.
Details are described in Section \ref{experiment3}.

\sh{Next, we evaluated our proposed methods on the standard benchmark following~\cite{chung2020defence}, where VoxCeleb2~\cite{chung2018voxceleb2} is used for training and VoxCeleb1~\cite{Nagrani17}, SITW~\cite{mclaren2016speakers}, and CNCeleb~\cite{li2022cn} are used for testing.
Since VoxCeleb2 contains speech from speakers spanning a wide range of different ethnicities, accents, professions, and ages recorded in various environments, we assume that there exists domain difference intrinsically within the dataset.
The development set of VoxCeleb2 contains $5,994$ speakers in total and is entirely disjoint from the test datasets.
VoxCeleb1 test set and SITW-Eval.Core are widely-used evaluation datasets
, and it is worth noting that the acoustic condition of them is similar with that of the training set VoxCeleb2, but there is domain difference intrinsically present between and within the datasets.
CNCeleb is a large-scale speaker dataset comprised of 11 different genres from 3,000 Chinese celebrities, and we utilize CNCeleb.E for testing.
The acoustic condition of CNCeleb is severely different from that of VoxCeleb.
}


\begin{figure}[]
\vspace{-15pt}
\centering
\includegraphics[scale=0.35]{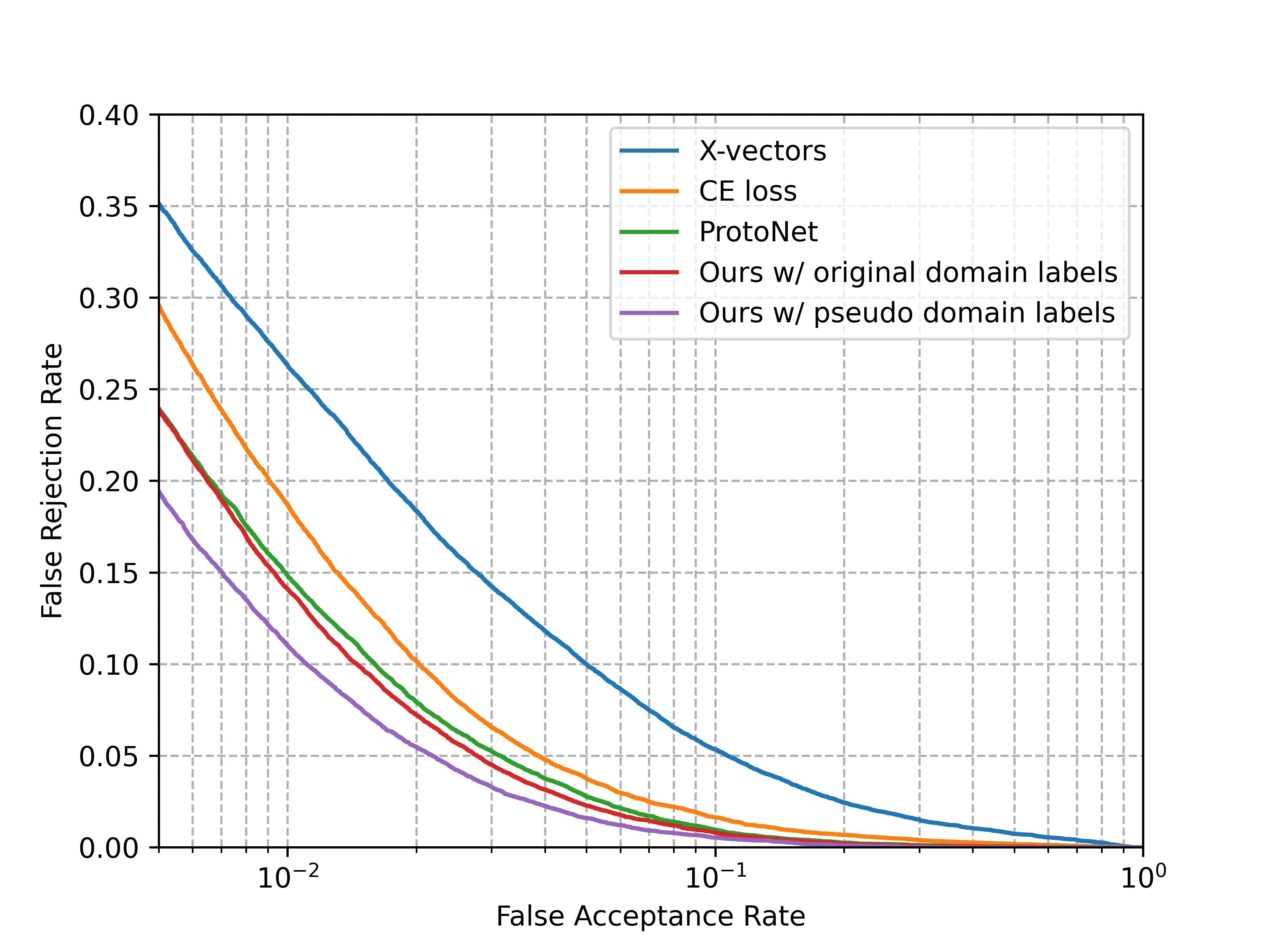}
\resizebox{0.9\columnwidth}{!}{
\begin{tabular}{c|cc}
\hline
Model                          & EER (\%)   & FRR @ FAR 10\% \\ \hline \hline
X-vectors                      & 5.685 & 3.817          \\ \hline
CE loss                        & 3.879 & 1.716          \\ \hline
ProtoNet                       & 3.496 & 1.257          \\ \hline
Ours w/ original domain labels & 3.231 & 1.016          \\ 
Ours w/ pseudo domain labels   & 2.717 & 0.756          \\ \hline
\end{tabular}}
\vspace{-5pt}
\caption{ROC curve, EER, and FRR for different methods.}
\vspace{-15pt}
\label{fig3}
\end{figure}

\vspace{-11pt}
{\bf Implementation details.} We use the x-vector system \cite{snyder2018xvec} and the speaker embedding network \cite{yun2019end} for evaluating \sh{on the artificially generated dataset.}
The x-vector system is a standard DNN speaker verification system using temporal statistics.
The speaker embedding network (SE network) uses an 1D-CNN architecture, and \cite{yun2019end} shows that SE network can effectively learn with short utterances.
We use both systems for the baseline, and we adapt our episodic learning algorithm to SE network.
We trained the network with $5$-way $5$-shot training episodes where 5 support and 5 query points per speaker are included.
Given 4 source domains (clean, babble, car, and music), we trained the domain-specific networks with two different setups: $N = 4$ (original) and $M = 3$ (clustered).
We chose the balancing parameters as $\lambda_{dg}=0.8$ and $\lambda_n=\{0.1, 0.2, 0.3\}$.

For evaluating our framework on the standard benchmark, we exploit Fast ResNet-34 with angular prototypical loss function~\cite{chung2020defence} as a baseline.
No data augmentation is performed during training, apart from the random sampling.
We trained the model with $200$-way $2$-shot training following~\cite{chung2020defence}.
We adapt our episodic learning algorithm to Fast ResNet-34 with $\lambda_{dg}=0.1$ using the same training strategy with the baseline for a fair comparison.
We adapt our clustering algorithm to assign pseudo domain labels to training samples and use them for training domain-specific networks and conduct ablation studies with various number of domains $M = 2,3,4$.

{\bf Evaluation metrics.} We used False Acceptance Rate (FAR) and False Rejection Rate (FRR).
FAR is the percentage of identification instances in which unauthorised persons (imposters) are incorrectly accepted, while FRR is that
in which the authorised person (target user) is incorrectly denied.
The Equal Error Rate (EER) is measured at which FAR and FRR are the same.
The minimum detection cost of the function (MinDCF) was defined by the NIST SRE evaluation plans~\cite{NIST_eval}. DCF is a weighted sum of FRR and FAR as $C_{fr} \cdot \text{FRR} \cdot P_{target} + C_{fa} \cdot \text{FAR} \cdot (1-P_{target})$.
In particular, the parameters $C_{fr} = 1$, $C_{fa} = 1$ and $P_{target} = 0.05$ are used for the cost function.

\vspace{-5pt}
\subsection{\sh{Comparison Studies on Generated Domains}}
\sh{On the artificially generated domains}, we compare our framework with the following methods: the x-vector system (X-vectors) and SE network trained with cross-entropy loss (CE loss) which do not consider 
both novel users and domains, the prototypical network (ProtoNet) which considers test case of novel users but not novel domains.

Fig. \ref{fig3} shows the ROC curve and the table of EER and FRR of the 5 different methods.
X-vectors are robust speaker embeddings given long utterances, but under short-utterance setting, where all utterances are cropped into 2-sec-long, SE network outperforms x-vectors.
For this reason, we 
applied our learning scheme to the SE network and showed 4 different results in Fig. \ref{fig3}.
Our method achieved the best performance compared with other learning frameworks.
It shows that our learning scheme can lead the network to extract discriminative features even for the unseen users and domains.
Surprisingly, our method w/ pseudo domain labels could achieve memory efficiency and produce less error compared with the original domain labels.
For better domain generalization, the domain-specific network should have distinct domain characteristics, {\it i.e.}, trained with well-separated domain features.
In our case, we define noise type as the domain, and some noise types could have similar characteristics in the feature space.
Thus, pseudo domain labels obtained with clustering performed better than the original domain labels. 
Also, this leads to the reduction in memory requirements for domain-specific networks. At 10\% FAR, ours w/ pseudo domain labels shows 55.94\%, 39.86\%, and 25.59\% relative FRR improvements compared to the other three methods using SE network. 


\begin{figure}[]
	\centering
	\includegraphics[width=1.01\columnwidth]{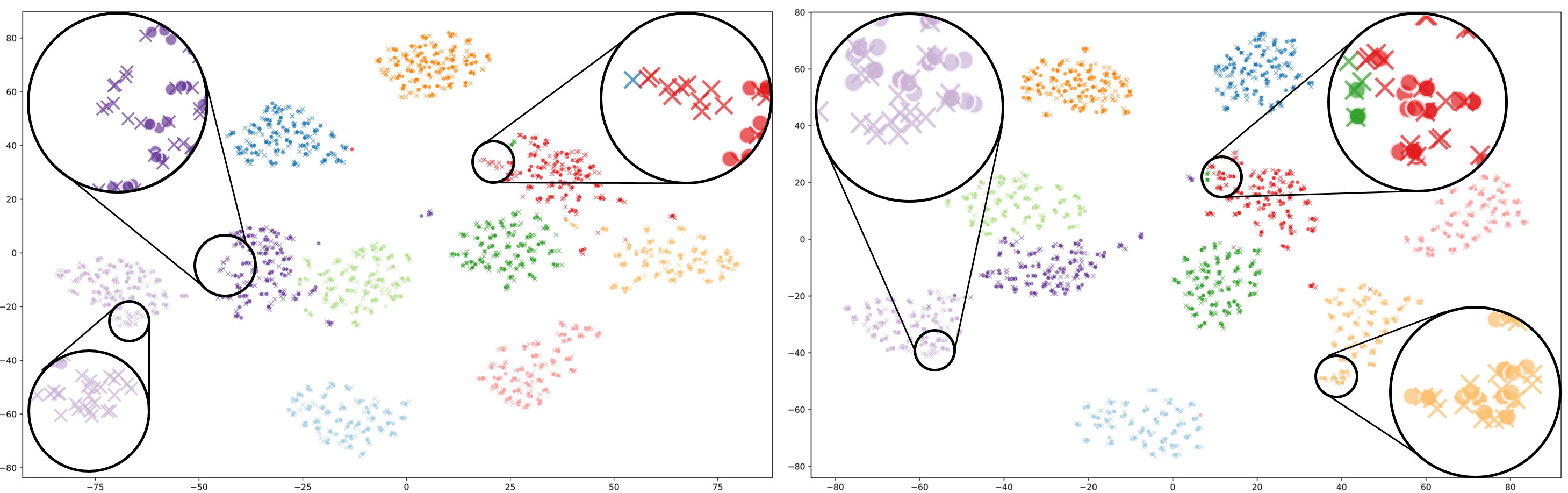}
	\vspace{-10pt}
	\caption{T-SNE plot of the features obtained by ProtoNet (left) and our framework (right). Color and shape represent speaker and domains, respectively. O and X indicate the feature of in-domain and out-domain, respectively.} 
\label{fig4}
\vspace{-5pt}
\end{figure}

\begin{table}[]
\caption{Average EER (\%) of four methods given in-domain and out-domain data with different numbers of domains.
}
\vspace{-5pt}
\centering
\resizebox{0.98\columnwidth}{!}{
\begin{tabular}{c|cccc}
\hline
Model                 & In-4 & Out-3 & Out-11 & Out-6 \\ \hline \hline
CE loss                        & 3.683      & 4.400  & 4.781             & 10.708      \\ \hline
ProtoNet                       & 2.888      & 3.649  & 4.055             & 9.707       \\ \hline
Ours w/ original domain labels & 2.895      & 3.417   & 3.873        & 9.265       \\
Ours w/ pseudo domain labels   & 2.528      & 2.916    & 3.396             & 7.185     \\ \hline
\end{tabular}}
\label{table3}
\vspace{-10pt}
\end{table}

\vspace{-5pt}
\subsection{Additional Analyses}
\label{experiment3}
{\bf Visualization.} As shown in the t-SNE plot~\cite{tsne} of embedding vectors in Fig. \ref{fig4}, our few-shot domain generalization framework extracts well-clustered features per each user in a domain-agnostic way.
Features extracted from ProtoNet are clustered well per each user but some features of out-domains are far from those of in-domains and the cluster center. The features could be easily rejected during verification, and it leads to higher FRR.

{\bf Performance on in-domains and out-domains.} In Table \ref{table3}, 
the first and second column respectively indicates the EER given the test data from in-domain (4 source domains) and out-domain (3 novel domains) data.
CE loss and ProtoNet use data augmentation to be robust to noise environments. However, they are vulnerable to novel domains that cannot be covered by noise augmentations.
Our method could adapt well to out-domains by explicitly considering the domain generalization test case during training.

{\bf Test on various target domains.}
We evaluated our framework using various out-domain settings: in Table \ref{table3}, the third and the fourth column shows EER with 11 novel noise types (Out-11) and 3 novel noise types with an additional severe SNR value (Out-6). Especially, in Out-6 case, we applied 2 different SNR values (-6, 0dB) to 3 novel noise types. Our framework shows better results especially for the Out-6 which has large domain gaps due to severe noise.
It demonstrates that the training scheme to deal with test case of unseen domains can lead the network to extract discriminative features even under the situation where train and test domains are severely different.

{\bf Train with various source domains.}
To study the effect of pseudo domain labels, we tested our framework with more source domains by augmenting original domains with 3 different SNR values (-6, 0, 6dB SNR). Overall, we have 10 domains: clean and 9 noisy domains.
Training 10 domain-specific networks requires much memory and compute, and we combine the original domains by clustering.
In Table \ref{table4}, ours w/ pseudo domain labels with $M=4$ that uses 4 domain-specific components achieved comparable performance compared to ours w/ original domain labels.
In addition, ours w/ pseudo domain labels with $M=10$ outperformed over ours w/ original domain labels, ensuring the efficacy of pseudo domain labels.

\vspace{-5pt}

\begin{table}[]
\caption{Effect of  number of pseudo domains on EER.} 
\centering
\vspace{-5pt}
\resizebox{0.98\columnwidth}{!}{
\begin{tabular}{c|c|cccc}
\hline
         & Original domain labels & \multicolumn{4}{c}{Pseudo domain labels} \\
 & $N$=10 & $M$=10        & $M$=8        & $M$=6       & $M$=4       \\ \hline \hline
EER (\%) &        2.264         &   2.044   &   2.145   &   2.176  &   2.332  \\ \hline
\end{tabular}}
\label{table4}
\vspace{-5pt}
\end{table}


\subsection{\sh{Evaluation on the Standard Benchmark}}
\sh{As shown in Table~\ref{table5}, our episodic learning strategy boosts the overall performance on all test datasets by a clear margin.
Our network learns the generalization ability from environments intrinsically present within VoxCeleb2. Hence, it works better for novel environments where the domain difference between training and test sets is not large. {\it i.e.}, VoxCeleb1 and SITW.
However, it is hard to improve the generalization ability on novel genres because VoxCeleb2 does not contain multiple genres, which indicates that our episodic learning setup cannot mimic the test scenario.
Thus, the performance marginally increases on CNCeleb, where the data are recorded for various genres, {\it e.g.}, singing, movie, drama, etc.
We expect that our episodic learning strategy on the training set of CNCeleb can improve the performance on these novel genres, 
which we leave for future experimental work.}

\begin{table}[]
\centering
\caption{EER and MinDCF of the baseline and ours with different number of pseudo domain labels (* is the number in~\cite{chung2020defence}).}
\vspace{-5pt}
\resizebox{1.0\columnwidth}{!}{
\begin{tabular}{c|cc|cc|cc}
\hline
\multirow{2}{*}{Model} & \multicolumn{2}{c|}{VoxCeleb1} & \multicolumn{2}{c|}{SITW} & \multicolumn{2}{c}{CNCeleb.E} \\
                       & EER          & MinDCF         & EER       & MinDCF       & EER          & MinDCF         \\ \hline \hline
Fast ResNet-34~\cite{chung2020defence} &   2.29 (*2.37)   &  0.188  &  4.02   &  0.301  &  15.83  &   0.815 \\ \hline
Ours w/ M=2         &  {\bf 2.09}  &  0.177 &  3.94  &   {\bf 0.278}  &   15.50   &  {\bf 0.717}   \\ 
Ours w/ M=3         &  2.12  &  {\bf 0.174} &  3.91  &   0.285  &   {\bf 15.44}   &  0.735   \\ 
Ours w/ M=4         &  2.16  &  0.177 &  {\bf 3.88}  &   0.280  &   16.04   &  0.727   \\  \hline
\end{tabular}}
\label{table5}
\vspace{-15pt}
\end{table}

\section{Conclusion}
We propose a domain generalization framework by generating two types of episodes to learn a speaker verification model that generalize to novel users and domains. We include domain clustering followed by learning domain-specific and -aggregation networks in source domain. The domain-aggregation network is learned to be domain-agnostic with episodes mimicking the test case of novel users and domains.
Extensive experiments show that our framework produces better domain-agnostic features and considerable performance improvements.

\bibliographystyle{IEEEtran}

\bibliography{references}

\end{document}